\documentclass[aps,pre]{revtex4}
\usepackage{graphicx,color}\begin{document}
\title{\bf{A model for the unidirectional motion of a  dynein molecule}}
\author{Sutapa Mukherji}
\affiliation{Department of Physics, Indian Institute of Technology, Kanpur
208016, India }
\date{\today}
\begin{abstract}
Cytoplasmic dyneins transport cellular organelles by moving on a microtubule 
filament. It has been found recently 
that depending on the applied force and the concentration of the 
adenosine triphosphate (ATP) molecules, 
dynein's step size varies. 
 Based on these studies, we propose a simple model
for dynein's unidirectional motion  taking  into account the variations 
in its step size.  We study how  the average velocity and
the relative dispersion in the displacement vary with the applied 
load. The model is amenable to further extensions by  inclusion of 
details associated with  the structure and the processivity of the 
molecule.

\end{abstract}
\maketitle
\section{introduction}
\label{sec:intro}
Dynein is a motor protein that moves on a  microtubule filament 
 to participate in certain activities  inside a cell
such as transport of vesicles, cell division etc \cite{alberts}. Dynein 
moves toward the negative end of the microtubule and this feature is used by 
the cell, e.g., for the transport or aggregation of  cellular organelles 
toward the nucleus of the cell.  
Based on their activities,  dynein molecules 
are classified into
two groups namely axonemal dynein and cytoplasmic dynein. Of these two,
the  axonemal dyneins  work together in  groups to  produce 
 rhythmic motions  of cilia and flagella. The cytoplasmic 
dynein, in which we are interested, perform,  
 either in group or alone, the 
job of   transport of cargoes such as cytoplasmic 
vesicles, chromosomes etc along the microtubule filaments. 
Besides these dyneins, there 
are other kinds of motor proteins such as kinesin and myosin
which are microtubule positive end directed and actin filament based 
respectively. These motor molecules also perform similar 
jobs as dynein. However, in comparison with these motors,
dynein molecules are exceptionally large and complex.
Roughly, the dynein molecule consists of two to three heavy chain units
each of which has a globular head domain (approximately $15$ nm diameter)
 with special sites capable 
of ATP hydrolysis, and two elongated structures known as stalk and 
tail. 
The stalk and the tail bind the microtubule and the cargo respectively. 
Because of the size, and also the difficulties in expressing
and purifying the mutants, the progress 
in understanding the mechanism of dynein's
motion  \cite{burgess,malliknature,toba} is slow. 

In general, the movement of a 
motor molecule takes place in the following
way. The adenosine triphosphate (ATP) molecule, 
which  diffuses in the solution inside the cell, 
can get attached to  specific regions or active sites in motor's  head. 
The motor molecule is capable of catalyzing  a 
 decomposition(hydrolysis) of this ATP  
into adenosine diphosphate (ADP), inorganic phosphate ($P_i$) and a
significant amount of energy.  
This energy so released 
causes a conformational change in the molecule resulting 
in a motion relative to the microtubule. This process of hydrolysis 
repeats if the ATP concentration  is sufficiently high   
and the motor molecule moves forward. In the  case of dynein, 
the head domain consists of  several sites at which ATP 
binding/unbinding or hydrolysis may take place. Although the 
details of the functional roles of these sites are still 
under investigation, it is believed that these sites 
are capable of controlling/regulating the motion of 
the dynein molecule.

Since the activities of motor proteins are force driven, it 
is important to know the response of the motion of the motor molecule
 to an externally applied 
force. The force generated by a motor molecule can be measured in 
single molecule experiments.  For example,  in 
optical trap experiments, a motor molecule is attached 
to a  polysterene bead which acts as a 
probe to monitor the movement of the motor molecule. 
The force on the bead due to the motor molecule 
is counteracted by a controlled backward force from the optical trap.  
The bead stays  immobile if the backward force is strong enough to 
resist the force generated by the motor.  
The minimum force needed to hold the bead immobile is the stall force 
which is a  measure of the 
force generated by the  molecule. 

An applied force, F,  can
affect the velocity in the following way.
In general, a motor molecule 
moves due to the internal conformational changes
 and these conformational changes occur due to the events such as  
ATP attachment or detachment and  ATP hydrolysis. 
The process of  ATP attachment is also 
dependent on the concentration of the 
ATP molecules inside the cell. 
These events take place with 
certain probabilities which determine their average rates.
The external force may   
give rise to  internal deformations  because of which the 
ATP attachment or detachment rate  and the  hydrolysis rate change 
thereby changing  the average velocity, $v$, of the molecule.
Thus, studying the  force-velocity relationship both 
theoretically and experimentally is crucial for predicting
how exactly various processes are affected by the force. 
In addition to this systematic force, there can be 
thermal noise which also affects the velocity of dynein  by 
changing  different rate constants. 
In case of such a stochastic  displacement of the molecule, 
it is possible to define an 
effective diffusion constant  
 $D_{\rm eff}=[\langle x^2(t)\rangle-\langle x(t)\rangle^2]/(2 t)$ 
which is a measure of the fluctuation in the displacement. 
Here, $\langle .\rangle$ denotes the statistical average.
In terms of this diffusion constant, one can also define 
a natural time scale  $a^2/D_{\rm eff}$ which is 
the time required for the molecule to diffuse a natural length 
scale, $a$. The distance traveled during this time due to the drift
velocity $v$ is $v\times a^2/D_{\rm eff}$. The ratio of these two 
distances is often referred as the randomness parameter 
\cite{schnitzer,svoboda}
\begin{eqnarray}
r=2 D_{\rm eff}/a v.\label{rand}
\end{eqnarray}   
This parameter derived earlier for kinesin 
has been found useful for understanding the 
internal mechanism of the molecule.

From  the structural studies of dynein, it is believed that  the 
dynein based transport in the cell is quite different and robust 
 in comparison with other motors.   
 However, experimentally, the mechanism of dynein's function
is not yet fully understood. For example, a recent experiment \cite{toba} 
that demonstrates $8$ nm step size for a cytoplasmic dynein  
contradicts an earlier observation \cite{malliknature} that
 the step size of dynein increases 
from $8$ nm to $16$, $24$ and $32$ nm  as the strength 
of the applied force reduces. Various step lengths found in 
\cite{malliknature} are  multiples of a unit 
step size $a=8$ nm, that corresponds to a certain periodicity of the 
microtubule filament. In addition, the results of \cite{toba}
on the average velocity of 
dynein and the value of the stall force are significantly  different 
from those of reference \cite{malliknature}. Single molecule experiment
on axonemal dyneins has also demonstrated step-wise displacements 
of $8$ nm size \cite{hirakawa}. In view of these differences, 
a theoretical model, that can take into account the wide step size 
variation of the molecule and can predict how, in this case,
 the velocity depends on the applied force, appears meaningful.
In addition, this analysis, which is going to be 
based on several assumptions on 
how various sites in dynein's head 
participate in regulating the motion, is expected to help 
explain the mysterious roles of these  sites.
    Although some theoretical work has been done
to understand various activities
of dynein such as 
rhythmic beating of axonemal dyneins \cite{elston2},  
unidirectional motion of  dyneins \cite{mallikpnas,ping},  
geared  motion of dynein to  
explain its force dependent step size \cite{cross}, 
so far there has been no
analytical approach to  understand 
 how the variable step size can affect the average velocity.

Based on the recent structural observation, we 
propose a model in which  one can incorporate $8$ nm  and other 
longer  jumps and study how this wide variation of the step size 
 affects the average velocity.
In case of variations in the step size, one needs to do 
a detailed fluctuation analysis to obtain the  distributions of the 
step-length and   the number of steps in a given time. 
The definition of the randomness parameter  in Eq. \ref{rand}, 
therefore, needs to be modified 
appropriately, possibly, by the inclusion of an average step length. 
However, in terms of $a$ as the smallest step, the randomness parameter 
helps us observe  how the step size variation 
affects the fluctuation in the displacement of the molecule.    
 Although, for simplicity, 
we include only a few essential structural and mechanochemical 
 details \cite{mallikpnas}, the model is flexible enough to 
incorporate other   details. 

The paper is organized as follows. In the next section, we discuss the basic 
assumptions based on which the model is built. The analysis of the model 
is presented in section \ref{sec:analysis}. This section is divided into two
subsections as we consider different variants of the model by including 
longer jumps of the dynein molecule. We conclude the paper with a summary 
of the work in section IV. 

\section{basic assumptions of the model} 
\label{sec:assumption}
It is known  that the dyenin's head domain consists of six sub-domains
of AAA family of proteins  arranged in the form of a ring around a 
central cavity
 \cite{samso}. Out of these AAA sub-domains, only four  
are capable of ATP binding with varying binding affinities 
\cite{reck,kon}. 
 ATPase activities   of these different AAA sub-domains are subjects
of experimental  investigations. 
 Although AAA1 is believed to be the primary site 
of ATP hydrolysis which powers the motion of the molecule, 
recent experimental observations support 
hydrolytic activities of other AAA sub-domains \cite{kon,takahashi,mogami}. 
Some of the studies elucidating the  functional roles 
of the AAA-sub-domains indicate that other  AAA sub-domains 
may play various regulatory roles by altering the hydrolysis
of the primary site and subsequently affecting the molecular function, 
or by modulating the efficiency of 
the coupling between the microtubule binding site and 
the primary site \cite{silv,reck,kon}. In our analysis, we call the 
sub-domains AAA2-AAA4 as secondary sites.

In view of the inherent complexity of dynein and our evolving 
understanding of how various components of the molecule function, 
it appears worthwhile to propose a model with the following assumptions.

1. For our entire analysis, we assume the dynein molecule to be 
effectively single 
headed \cite{mallikpnas}. 
Studies on head-head coordination \cite{shima} 
during the processive movement of dynein suggest that 
similar to kinesin and myosin, 
the rear head of dynein may sense a pulling force 
from the  forward head and this  force  may  
modulate the kinetic steps appropriately to coordinate the forward motion. 
The hydrolysis presumably takes place alternatively 
in each head and at a given step a single head 
proceeds forward in a hand-over-hand fashion. Since these details 
of the head coordination is not an issue here, both the heads can be 
thought of as a composite object - a single head. 
The effect of the pulling force in the ATPase activities of the head is 
assumed to be incorporated through the rates.   
The two-head problem can be studied if additional information is known
about how the force exactly modulates the kinetic steps in the head.

2.  We assume that there  are only two ATP binding sites (sub-domains) 
of which one is a primary site (P) and the other one is a secondary site (S).
This is modified later by including more number of secondary sites.

3. In our analysis, only the primary site is  capable of ATP hydrolysis. 
It is possible to incorporate hydrolytic activities of other secondary 
sites  
considering that such additional hydrolysis may change the driving 
force or the regulatory mechanism of dynein. We plan to consider 
these  processes in our future analysis elsewhere. 

4. Rates of binding or unbinding of ATP  to or from the primary  site
are denoted by $k_{\rm on,1}$, $k_{\rm off,1}$ respectively.  
Subscripts 2 and 3 
with these rates imply binding or unbinding  processes at the 
secondary sites.
 It is assumed that the primary site has the highest affinity
to the ATP molecule. Therefore, if the primary site is empty, it first
 gets filled even in the presence of an empty secondary site.
Similarly, ATP unbinding from the  primary site does not take place 
if there are occupied secondary  sites. These rules hold  throughout 
the entire analysis.

5. To start with, it is assumed that  ATP hydrolysis at 
the primary site causes
the dynein to jump the smallest distance $a=8$ nm. 
This can be modified further.
The distance,  a dynein molecule can jump after a hydrolysis
 can change depending on whether the secondary site is occupied or
unoccupied. If the secondary site is occupied, the dynein molecule
jumps a distance $a$ after a hydrolysis. In case the secondary site
is unoccupied, the dynein molecule moves $2 a$ distance ahead at a time 
after a hydrolysis. In the presence of two secondary sites, 
the dynein can jump 
$3a$ distance at a time 
 after one hydrolysis if both secondary sites are empty.
Jumps of length $2a$ or $a$  are possible if one or both the 
secondary sites are occupied respectively. The hydrolysis processes, 
that trigger
jumps of length $a$, $2a$ or $3a$, take place at rates $k_{\rm cat,1}$,
$k_{\rm cat,2}$ and $k_{\rm cat,3}$ respectively. A  schematic diagram, 
indicating  various moves described here, is presented in figure 1. 

\begin{figure}[htbp]
   \includegraphics[width=3in,clip]
{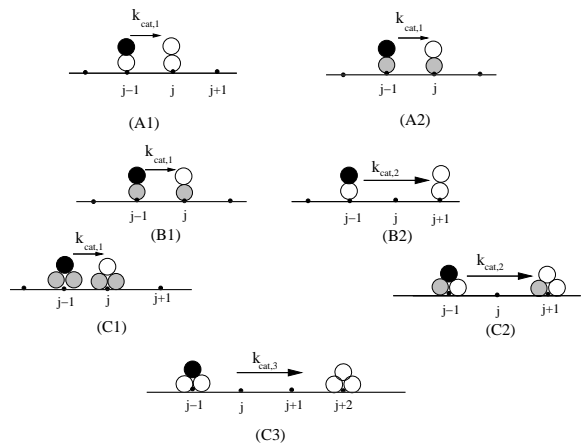}
    \caption{Various moves of the dynein molecule, considered in the 
text, are shown here.
AAA sub-domains or ATPase activity sites 
are represented by circles. The filled, shaded and unfilled 
circles represent ATP occupied primary site, ATP occupied secondary site 
and and empty primary or secondary site, respectively. Various
 moves take place
after hydrolysis in the primary site with the rates as indicated in 
Eq. \ref{catrate}. 
(A) A jump by a distance $a$ always. The dynein molecule is 
assumed to have two active AAA sub-domains. 
  (B) The molecule jumps a distance 
$a$ (or $2a$) when the secondary site is in an ATP-bound(or unbound) state.
The dynein molecule has two active AAA-sub-domains. This case 
 has been discussed in section \ref{sec:twostep}. 
(C) Three 
possible moves of lengths $a$, $2a$ and $3a$ depending on the number of 
ATP-bound secondary sites. The dynein molecule has three active AAA 
sub-domains. This case has been discussed in section \ref{sec:threesite}.}  
\label{fig:pathway}
\end{figure}

6. Our analysis is based on the  assumption that the time required for 
a transition is much smaller than the time between two 
successive transitions. In addition, we also assume that 
 the thermal relaxation of the molecule after a transition is much 
faster compared to the transition rate \cite{leibler,fisher}. This allows us 
to assume quasi-equilibrium  for transitions from one state to the 
other  over various free energy barriers. 
At temperature T, the rate constants are, therefore,
 expected to depend  exponentially on  
$[F D/(k_B T)]$, where $D$ is an 
appropriate parameter in the units of length and $k_B$ is the Boltzmann 
constant. Our next assumption is related to the explicit dependences
of various rate constants on the load.  

7. It is expected that the mechanochemical cycle is affected by applied load,
in particular, since 
we know that the motion of the motor comes to a halt when the 
applied opposing force is sufficiently high. One way of implementing this 
load dependence is by considering a load dependent hydrolysis rate as  
\begin{eqnarray}
k_{\rm cat,i}=A(s) k_{\rm cat,0}
\exp[-\alpha F d(s)/k_B T],\label{catrate}
\end{eqnarray} where $i=1,\ 2,\ 3$ and   $d(s)=i\times a$
is the distance, the  molecule jumps after the hydrolysis.
 $k_{\rm cat,0}$ is the hydrolysis
rate for no load. $\alpha$ is the load distribution factor for
hydrolysis. In general, there is no restriction on the 
sign of $\alpha$. Here, it is more intuitive to assume $\alpha$ 
to be positive since the hydrolysis rate should decrease with the increase 
in the opposing external force. There exists, however, one restriction on 
$\alpha$ that the sum of the load distribution factors for hydrolysis 
and reverse hydrolysis should be unity.   The possibilities 
of reverse hydrolysis is not taken into account in order to appreciate 
the crucial features of the results when the number of parameters
is less.
ATP hydrolysis is assumed to be enhanced if at least one
secondary site binds ATP. This is taken care of by $A(s)$ which is $1$ if the
secondary site is ATP-bound and is equal to $0.01$ otherwise. In case of 
jumps by only a distance $a$, 
we have only one hydrolysis rate $k_{\rm cat,1}$ 
 irrespective of whether
the secondary site is occupied or unoccupied by an ATP molecule.
  In this case, $A(s)=1$ always. In case of jumps by distance $a$ or 
$2a$, we choose 
the values of $A(s)$ as prescribed here. 

As has been proposed earlier \cite{mallikpnas}, 
we also assume that the ATP binding affinities of the 
secondary sites are dependent on the load as
\begin{eqnarray}k_{\rm on,2-4}=
k_{\rm on,2-4}(F=0) \exp[F d_0/k_B T],\label{kon2}
\label{bindaff}
\end{eqnarray} where $d_0$ is an adjustable parameter in the 
units of length. For obtaining all the results,  we use the value 
 $d_0$ same as that  in reference \cite{mallikpnas}.

\section{analysis of the model}
\label{sec:analysis}
The forward motion can be described through  equations that describe
time evolutions of certain  variables on a one-dimensional lattice. 
A  variable, here,  represents the 
probability of a molecule being at a given lattice-site at a time $t$ 
with a given 
configuration of its ATPase activity sites. For example, a  
variable $S_j^{\alpha \beta}$ represents the probability of a molecule 
with two AAA sites, being at the $j$th lattice-site with 
primary and secondary  sites being in 
$\alpha$ and $\beta$ states respectively.  $\alpha$ and $\beta$ 
can have values $0$ or $1$ if a site is unoccupied or occupied 
by an ATP. The time evolution equations for all the  variables, say, 
$S_j^{00},\ S_j^{01},\ S_j^{10},\ S_j^{11}$ for a molecule with two ATPase
sites,  can be combined into a matrix equation. 
This matrix equation, thus, represents the time evolution 
of the column matrix $\rho_j$ whose elements are the probability
variables mentioned above. 
This equation can further be recast as a matrix equation 
describing the time evolution of the column matrix 
$\sum_{j=-\infty}^{j=\infty} \zeta^j \rho_j$.  $\zeta$ is an 
arbitrary parameter and as we shall discuss below, the determination
 of the average velocity requires $\zeta$ to be unity. 

In the following subsection, we discuss the specific case  where the 
molecule jumps by a distance $a$ only irrespective of the ATP occupancy 
state of the secondary site. The derivation of the velocity is presented
here in detail. This method allows  us to 
 look at the variation of the randomness parameter with 
the applied opposing force. The same method is generalized in the next 
subsections for the cases where the molecule can make longer jumps 
depending on the ATP occupancy state of the secondary sites.  

\subsection{Dynein with one primary and one secondary site}
 \label{sec:twosite}
\subsubsection{Jumps by a  distance $a$}
\label{sec:onestep}
The time evolution of the four probability variables
corresponding to all four configurations of the ATP binding
sites of a dynein at the $j$th site is governed by the equations 
\begin{eqnarray}
&&\frac{d S_j^{00}}{dt}=k_{\rm off,1} S_j^{10}+k_{\rm off,2} S_j^{01}+
k_{\rm cat,1} S_{j-1}^{10}-k_{\rm on,1} S_j^{00}\\
&&\frac{dS_j^{01}}{dt}=k_{\rm cat,1} S_{j-1}^{11}-k_{\rm on,1}
S_j^{01}-k_{\rm off,2} S_j^{01}\\
&&\frac{dS_j^{10}}{dt}=k_{\rm off,2} S_j^{11}+k_{\rm on,1} S_j^{00}-
k_{\rm off,1} S_j^{10}-k_{\rm on,2} S_j^{10}-k_{\rm cat,1} S_j^{10}\\
&&\frac{dS_j^{11}}{dt}=k_{\rm on,1} S_j^{01}+k_{\rm on,2} S_j^{10}-
k_{\rm cat,1} S_j^{11}-k_{\rm off,2} S_j^{11}.
\end{eqnarray}
These equations can be written in a matrix form as
\begin{eqnarray}
\frac{d\left[{\bf \rho_j}\right]}{dt}=
\left[{\bf A}\right]\left[{\bf \rho_j}\right]+
\left[{\bf B}\right]\left[{\bf \rho_{j-1}}\right],\label{maineqn1}
\end{eqnarray} where
\begin{eqnarray}
{\bf \rho_j}=\left(\begin{array}{cc} S_j^{00}\\S_j^{01}\\S_j^{10}\\S_j^{11}
\end{array}\right),
\end{eqnarray}
and ${\bf A}$ and ${\bf B}$ are matrices whose elements are the  
various rate constants of the differential equations. 
 Multiplying both sides of Eq. \ref{maineqn1} with
$\zeta^j$ and summing over all possible values of $j$, we have
\begin{eqnarray}
\frac{d}{dt}\sum_{j=-\infty}^\infty \zeta^j\rho_j=
\left[{\bf A}\right]\sum_{j=-\infty}^\infty \zeta^j\rho_j+\left[{\bf
B}\right]\sum_{j=
-\infty}^\infty \zeta^j\rho_{j-1}.
\end{eqnarray}
In terms of  $G(\zeta,t)=\sum_{j=-\infty}^\infty \zeta^j\rho_j$,
the above equation is
\begin{eqnarray}
\frac{d}{dt} G(\zeta,t)=\left(\left[{\bf A}\right]+
\zeta\left[{\bf B}\right]\right)  G(\zeta,t)=
\left[{\bf R}(\zeta)\right] G(\zeta,t),
\end{eqnarray}
where
\begin{eqnarray}
\left[{\bf R}(\zeta)\right]=
\left(\begin{array}{llcl} -k_{\rm on,1}
 \ \ \ \ \ \ \ \ \ \ \  k_{\rm off,2}\ \ \ \ \ \  \ \ \ \ 
\ \ \ \ k_{\rm off,1}+\zeta k_{\rm cat,1}\ \ \ \ \ \ \ \ \ \ \ \ \ \ \
  0\\0\ \ \ \ \ \ \ \ \ \ -(k_{\rm on,1}+k_{\rm off,2})
\ \  \ \ \ \ \ \ \ \ \ 0\ \ \ \ \ \ \ \ \  \ \ \ \ \ \ \ \  \ \ \ 
\ \ \ \ \zeta k_{\rm cat,1}\\
k_{\rm on,1}\ \  \ \ \  \ \ \ \ \ \ \ \ \ \ 0\ \ \ \ \ 
 \ \ \  -(k_{\rm off,1}+k_{\rm on,2}+k_{\rm cat,1})\ \ \ \ \ \ \ \ \
 k_{\rm off,2}\\
 0\ \  \ \ \ \ \ \ \ \ \ \ \ \ \ \ \ 
k_{\rm on,1}\ \ \ \ \ \ \ \ \ \ \ \ \ \ \  \ \ k_{\rm on,2}\ \
\ \ \ \ \ \ \ \ \ -(k_{\rm cat,1}+k_{\rm off,2})
\end{array}\right)
\end{eqnarray}
$\left[{\bf R}(\zeta=1)\right]$ is a transition matrix with the sum of
all elements in a column being zero. The largest eigenvalue of the matrix
${\bf R}(\zeta=1)$ is, therefore, zero.

Our final aim is to find out the average velocity $\langle j\rangle/t$
and the diffusion constant
of the dynein molecule. It has been shown earlier that the average velocity
 and the effective diffusion constant
in such a case can be found from the relations \cite{elston,mogilner}
\begin{eqnarray}
\langle v\rangle =a \frac{\langle j\rangle}{t}=a\lambda_l'(1),\ \ \
D_{\rm eff}=\frac{a^2}{2}(\lambda_l''(1)+\lambda_l'(1)),
\end{eqnarray}
where $\lambda_l(\zeta)$ is the largest eigenvalue of
$\left[{\bf R}(\zeta)\right]$ and the primes denote derivatives of
$\lambda$ with respect to $\zeta$.

In order to find out the largest eigenvalue,
it is necessary to solve the
 characteristic equation,
\begin{eqnarray}
Det [{\bf R}(\zeta)-\lambda {\bf I}]=0.\label{chareqn}
\end{eqnarray}
The characteristic equation  can be solved for the eigenvalues
straight away. However, in order to find the derivative of the largest
eigenvalue at $\zeta=1$, it is convenient to  substitute $\zeta=1+\delta$,
and
$\lambda=\delta\lambda'(1)+\frac{\delta^2}{2} \lambda''(1)$ in Eq.
\ref{chareqn}, and then find out $\lambda_l'(1)$ and
 $\lambda''_l(1)$ by
equating coefficients of $\delta$ and $\delta^2$,
respectively,  to zero.
The average velocity found this way is 
\begin{eqnarray}
&&v=a\times\nonumber\\
&&\frac{k_{\rm cat,1} k_{\rm on,1}[(k_{\rm cat,1} k_{\rm off,2}+
(k_{\rm off,2}) k_2+
k_{\rm on,1} k_2]}
{ (k_{\rm cat,1})^2 k_{\rm off,2}+k_{\rm cat,1}(k+
k_{\rm off,2} k_2+2 k_{\rm off,2} k_{\rm on,1})+(k_{\rm off,2}+
k_{\rm on,1})(k_{\rm off,1} k_{\rm off,2}+
k_{\rm on,1}k_2)},\label{velocity}
\end{eqnarray}
where $a$ is the step size, 
$k=k_{\rm off,1}k_{\rm off,2}+k_{\rm on,1}k_{\rm on,2}$ and 
$k_2=k_{\rm off,2}+k_{\rm on,2}$.
The expression for the diffusion coefficient can be obtained in
the similar way. Since the diffusion coefficient
involves a more complicated algebraic dependence on various rates, we 
avoid mentioning the expression here. Instead, we focus on the variation 
of the randomness parameter with the applied force later.   

\vskip .3in
\begin{center}
Table 1\\
Model parameters\\
\end{center}
\vskip .3in
\begin{tabular}{|c|c|c|}\hline
Symbol & Value & Meaning\\
\hline
$k_B T$ & $4.1\ pN \ nm$ & Thermal energy\\
$k_{\rm off,1}$ & $10 \  s^{-1}$ & Rate of unbinding of an ATP molecule from 
the primary site\\
$k_{\rm off, 2}$ & $250 \ s^{-1}$ & Rate of unbinding of the first   ATP 
molecule from 
one of the  secondary  sites\\
$k_{\rm off,3}$ & $250 \ s^{-1}$ & Rate of unbinding of the second ATP 
molecule from one of the secondary sites\\  
$k_{\rm on, 1}$ &  $4\times 10^5 M^{-1} s^{-1}\ [ATP]$ & 
Rate of ATP binding to 
the primary site\\
$k_{\rm on,2}(F=0)$ & $4 \times 10^5 M^{-1} s^{-1} [ATP]$& Rate of  ATP 
binding at the secondary sites under zero load\\
$k_{\rm on, 3}(F=0)$ & $k_{\rm on,2}(F=0)/4$ &  Rate of second ATP binding 
at the secondary site\\ 
$d_0$ & $6\ nm$ & an adjustable length as introduced 
in Eq. \ref{kon2}\\
$k_{\rm cat,0}$ & $55 \ s^{-1}$ & hydrolysis rate under zero load\\
$\alpha$ & $0.3$ & Load distribution factor\\
\hline
\end{tabular}

\vskip .5in

\begin{figure}[htbp]
   \includegraphics[width=3in,clip]
{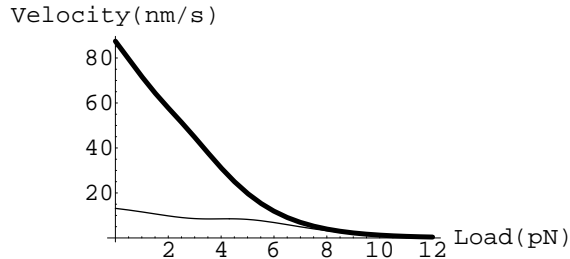}
    \caption{A plot of the velocity $(nm \ s^{-1})$ with force $(pN)$ 
for different ATP concentrations. The molecule can move forward 
by jumping a distance $a$ only . The thin and thick  lines correspond 
to $[ATP]=5\mu M,\ \  40\mu M$ respectively.}
\label{fig:forcevel1}
\end{figure}

\begin{figure}[htbp]
   \includegraphics[width=3in,clip]
{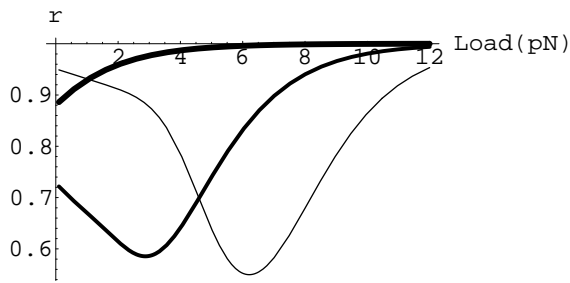}
    \caption{A plot of the randomness parameter, $r$, 
 with the applied force $(pN)$ at
various ATP concentrations. In this case the molecule can move forward 
by a distance $a$ only. Three  solid lines with increasing 
thickness correspond to 
ATP concentrations, $[ATP]=5 \mu M,\ 40\mu M,\ 2mM$ respectively.}
\label{fig:randpara1}
\end{figure}

Using  the values of  various model parameters as listed in table 1, we 
plot the force-velocity curve for two different ATP concentrations 
in  figure 
\ref{fig:forcevel1}. Since the hydrolysis rate decreases with the force, 
the velocity is ultimately 
expected to decrease with the force when force is large. It can be seen from 
Eq. \ref{velocity}, that the velocity decreases exponentially 
as $v\sim k_{\rm cat,1}=A(s) k_{\rm cat,0}  \exp[-\alpha F d(s)/k_BT]$ for 
high force. For moderate force, there is a possibility of an increase in the 
velocity. This is due to $k_{\rm on,2}$ which increases with the force 
as per Eq. \ref{kon2} and enhances the probability 
of the molecule being 
in the state $S_j^{11}$. The velocity increases as a result of hydrolysis 
in this state. This can be also verified by choosing $k_{\rm on, 2}$ 
to be independent of the force in which case the velocity curve decreases  
monotonically for all values of force. For higher values of ATP concentration,
the maximum in the velocity becomes less pronounced and it eventually 
disappears for high ATP concentration. The increase in the velocity at
low ATP concentration is  not present
in the experimental observations of \cite{toba}. However, it may not be 
appropriate to compare these results with the experiments of reference 
\cite{toba} 
 mainly because of the simplistic approach of the model 
where  dynein has been assumed to be single-headed with 
less number of secondary sites and  
with a very simple kinetic cycle without any reversal of hydrolysis 
or other additional hydrolysis.   
As we shall show later, the velocity changes significantly  
in the low force regime,   
as the   possibility of longer jumps is incorporated.
The variation of the randomness parameter $r$
with the force, for  given concentrations of ATP molecules, appears as in 
 figure \ref{fig:randpara1}. 
At low ATP concentrations, the 
 randomness parameter initially decreases with the force in the 
moderate force regime. As the force is increased further, the 
randomness parameter increases and approaches $1$.
 The increase in the velocity around the same values 
of force, seems to be a reason  for the  decrease in $r$ 
for  low concentrations of ATP.

\subsubsection{Possibility of jumps by a distance $2a$}
\label{sec:twostep}
\begin{figure}[htbp]
   \includegraphics[width=4.0in,clip]
{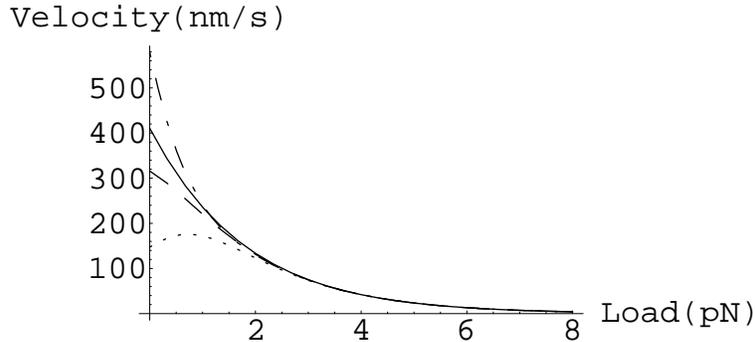}
    \caption{A plot of the velocity with load at
ATP concentration $[ATP]=2mM$. The thin solid line and the 
dashed line  correspond to the one-step
and two-step cases respectively. 
The dash-dot line represents the average velocity for three-step jump.
The dotted line represents the average velocity of
 the molecule with possibilities of three-step jump 
with $k_{\rm cat,2}$ reduced to 
$k_{\rm cat,2}=0.55 \exp[-\frac{0.3 \times 16\times F}{4.1}]$.} 
\label{fig:forcevel3}
\end{figure}

In the following, we incorporate the possibility of  jumps by a distance
 $2a$. The
dynein molecule moves forward by a distance $a$ or $2a$ after a hydrolysis
if the secondary site is in ATP bound or unbound state respectively 
(see figure 1B). 
Eq. \ref{catrate}
shows that the hydrolysis rate, $k_{\rm cat,1}$, that triggers
a shorter jump,  needs to be distinguished from
the hydrolysis rate, $k_{\rm cat,2}$,  which powers a  longer jump. 
Less binding affinity of the secondary site at a low force ensures that
there can be larger jumps that are  triggered by a hydrolysis whose rate
increases as the value of the force decreases. Thus,  we expect  a 
significant change in the average velocity in the low force region 
from that of  the case where the molecule jumps only by a distance $a$.  
The time evolution of the four probability
variables are given as
\begin{eqnarray}
&&\frac{d S_j^{00}}{dt}=
k_{\rm off,1} S_j^{10}+k_{\rm off,2} S_j^{01}+
k_{\rm cat,2}
S_{j-2}^{10}-k_{\rm on,1} S_j^{00}\\
&&\frac{dS_j^{01}}{dt}=k_{\rm cat,1} S_{j-1}^{11}-
k_{\rm on,1} S_j^{01}-k_{\rm off,2} S_j^{01}\\
&&\frac{dS_j^{10}}{dt}=k_{\rm off,2} S_j^{11}+k_{\rm on,1} S_j^{00}-
k_{\rm off,1} S_j^{10}-k_{\rm on,2} S_j^{10}-k_{\rm cat,2} S_j^{10}\\
&&\frac{dS_j^{11}}{dt}=k_{\rm on,1} S_j^{01}+k_{\rm on,2} S_j^{10}-
k_{\rm cat,1} S_j^{11}-k_{\rm off,2} S_j^{11}.
\end{eqnarray}
Proceeding in the similar way as the one-step case, we may combine the 
differential equations in a single matrix equation 
\begin{eqnarray}
\frac{d}{dt} G(\zeta,t)=\left[{\bf R}(\zeta)\right] G(\zeta,t),
\end{eqnarray}
where
\begin{eqnarray}
\left[{\bf R}(\zeta)\right]=
\left(\begin{array}{llcl} -k_{\rm on,1}
 \ \ \ \ \ \ \ \ \ \ \ k_{\rm off,2}
\ \ \ \ \ \ \ \ \ \ \ \  \ \ \ \ k_{\rm off,1}+\zeta^2 k_{\rm cat,2}
\ \  \ \ \ \ \ \ \ \  \ \ \ \ \ \ \ 0\\
0\ \ \ \  \ \ \ \ \ \  -(k_{\rm on,1}+k_{\rm off,2})
\ \  \  \ \ \ \ \ \  \ \ \ \ \ \ \ 0\ \ \  \ \ \ \ \ \ \ \ \ \ \ \ \
\ \ \ \ \   \ \ \ \zeta k_{\rm cat,1}\\
 k_{\rm on,1}\ \  \ \ \ \ \ \ \ \ \ \ \ \  0\ \ \ \ \ \ \  
\ \ \ \ \ \ \ \  -(k_{\rm off,1}+k_{\rm on,2}+k_{\rm cat,2})\ \
\ \ \ \ \ \  k_{\rm off,2}\\
 0\ \ \ \ \ \ \ \ \ \ \ \ \ \ \ \ \ \ k_{\rm on,1}\ \ \ \ \ \ \ \ \ \ \
\ \ \ \ \ \ \ \ \ \ \ \ \ \  k_{\rm on,2}
\ \ \ \ \ \ \ \ \ \ \   
-(k_{\rm cat,1}+k_{\rm off,2})
\end{array}\right)
\end{eqnarray}
and $G(\zeta,t)$ is the same as that defined earlier. 
The procedure for finding out the
 average velocity is the same as before.
Figure \ref{fig:forcevel3} shows a comparison of velocities for  pure
single step jump  and for the case where both one and two-step jumps
are possible.

\begin{figure}[htbp]
   \includegraphics[width=5in,clip]
{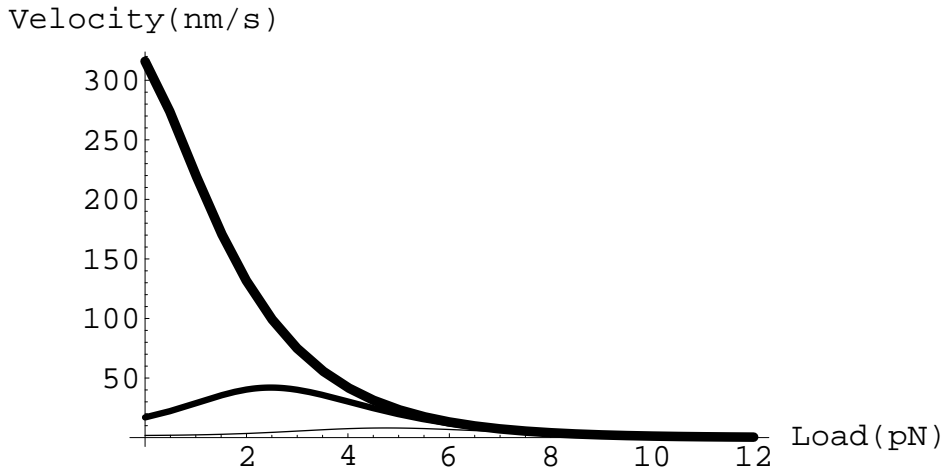}
    \caption{Force-velocity plot for  ATP concentrations 
$5\mu M$, $40\mu M$ and $2mM$. 
Lines with increasing thickness correspond to higher ATP concentrations.
The molecule is allowed to jump by a distance $2a$ at a time
 if the secondary site is 
empty.} 
\label{fig:forcevel2new}
\end{figure}

\begin{figure}[htbp]
   \includegraphics[width=3in,clip]
{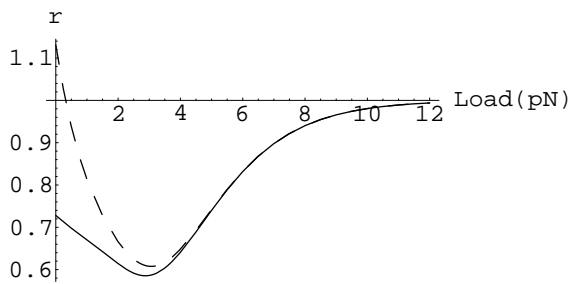}
    \caption{A plot of the randomness parameter with load at
$[ATP]=40\mu M$. The solid and the dashed  line correspond 
to the case 
where the molecule can jump a distance $a$ or $2a$ respectively.}
\label{fig:rand-para-compare}
\end{figure}

The  figure shows that the 
average velocity in two-step case reduces when the force is small.
 In case of jump by a distance $a$ only, the forward move takes place at a 
much higher rate. Since in  the two-step case, single steps are not allowed 
when the secondary site is empty  and  the allowed bigger 
moves take place at a much lower rate, the average velocity 
reduces. The force-velocity plots for different ATP concentrations 
have been shown in  figure \ref{fig:forcevel2new}.
Introducing the possibility of two-step jumps increases the fluctuation 
in the displacement at low forces (see figure \ref{fig:rand-para-compare}). 
This is likely to be the case since 
for low force, two-step jumps are more probable than for 
high value of force. At high forces, the probability of longer 
steps decreases and the two plots merge. 

\subsection{Dynein with one primary and two secondary sites: possibility 
of jumps by a distance $3a$}
\label{sec:threesite}
As further improvement toward the more realistic picture, we include
two secondary sites.
With three sites, there are now possibilities of jumps by 
a distance  $3a$ in one move (See figure 1C). 
This happens if no secondary site is occupied.
With two secondary sites, the dynein molecule at  $j$ th site
 can remain in 8 possible states with probabilities
$S^{000}_j,\ S^{001}_j,\ S^{010}_j,\ S^{100}_j,\
S^{011}_j,\ S^{101}_j, S^{110}_j, \ S^{111}_j$, evolution of
 which can be written in a way similar to that 
mentioned before.
The average velocity calculated for this case is shown in figure 
\ref{fig:forcevel3} along with the previous results for comparison.
The average velocity is quite sensitive to the hydrolysis rates. 
It can be seen that lowering $k_{\rm cat,2}$ to 
$k_{\rm cat,2}=0.55 \exp[-\frac{0.3 \times 16\times F}{4.1}]$,
causes a significant  decrease 
in the average velocity.

\begin{figure}[htbp]
   \includegraphics[width=5in,clip]
{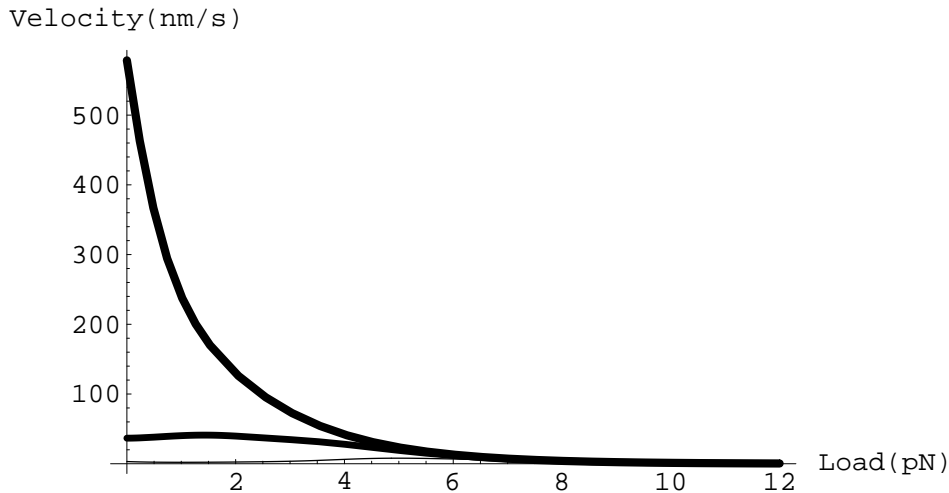}
    \caption{A plot of velocity with load at different ATP concentrations.
Three solid lines with increasing thickness correspond to
[ATP]= $5\mu M$, $40 \mu M$
and $2m M$. The molecule is allowed to jump a distance $3a$ at a time if 
no secondary site is in ATP-bound state. }
\label{fig:forcevel4}
\end{figure}

The concentration of ATP has  significant effects on the average velocity.
This is clear from the force-velocity plot in figure \ref{fig:forcevel4} 
for different ATP concentrations. 

\section{Summary}
\label{sec:summary}
Dynein's structure is known to be fundamentally 
different from other motor proteins such as kinesin and myosin. 
 It is believed that the complexity in the structure 
of dynein  leads to a robust  motion 
and leaves more opportunity of regulation of the motion
 at multiple levels. Due to its complexity, our knowledge about 
how dynein functions has been limited.  
One of the single molecule experiments suggests that unlike 
other motor proteins, dynein has a wide step size variation and the 
step length of dynein can be $8$, $16$, $24$ or $32$ nm depending 
on the strength of the opposing force. It has been predicted that the step 
size is around $32$ nm for low or no load situations and the step size 
decreases as the strength of the force increases. 
One of the more recent 
experiments, however, demonstrates a fixed, 
load-independent step size close to 
$8$ nm and different results regarding 
the value of the stall force and its dependence on the ATP 
concentration. 

Since  these experiments give different views about the step length 
and other properties of dynein, we propose our theoretical model 
to see the effect of  step size variation on 
the average velocity of the molecule and  
the fluctuation in its displacement.  Our model is based on certain 
simplifications. We assume that the  head domain has 
 a primary site which is primarily responsible for 
ATP hydrolysis that drives the motion and 
a maximum of  two secondary sites which can 
regulate the motion through ATP binding. The force induced 
reduction of the step size is introduced through the assumption that   
the opposing force increases the binding affinity of the secondary site
 and the molecule makes a shorter jump if more number of secondary sites 
are in ATP bound states.
Through this model, it is possible to understand analytically how the 
velocity changes as longer jumps are introduced  in dynein's motion  
at a given ATP concentration. At low ATP, 
we find an initial increase in the velocity  with the force before 
it finally decreases.  This increase in the velocity is not seen 
at high ATP concentrations. It appears that this increase 
is partly due to the force dependent binding affinity 
of the secondary sites. We also study the randomness parameter,
introduced in Eq. \ref{rand}. This quantity gives an 
estimate of the variation of the fluctuation in the displacement
with the applied opposing force.
The possibility of larger jumps under low-load 
condition, introduces significant changes  in the randomness 
parameter in the low-force regime.

The model presented here 
neglects certain features such as the presence of two heads of dynein
instead of  only one as it is presently simplified to, 
three secondary sites capable of rendering further drive or regulation 
through additional hydrolysis or ATP binding and the possibility of 
reverse hydrolysis. 
The mechanochemical cycle
is also oversimplified in our model. Despite these, we believe 
that predictions from our analysis can be 
tested through carefully designed experiments as they would provide an 
an indirect verification 
of the crucial assumptions set in the model such as the increase in 
ATP  binding affinity of the secondary sites 
with increasing load, dependence of the step-size of the molecule
on the number of ATP bound secondary sites. 
The analytical approach 
gives  explicit expressions of the velocity and 
the fluctuation in the displacement  in terms of 
various rate constants and one can also isolate the contributions
of different secondary sites.  We, therefore, believe that 
the use of site directed mutants that specifically inhibit 
the ATPase activity of different AAA sub-domains, together with 
other biochemical tools and single molecule experiments may be required 
to isolate the contributions of the secondary sites. The experimental 
knowledge thus generated regarding the roles of different secondary sites 
in powering or regulating the motion of the molecule will make it 
possible to further modify the model suitably. 
The model presented here is a first 
phenomenological step 
toward understanding the complete molecular model of dynein.


\end{document}